\documentclass[reprint,amsmath,amssymb,pra]{revtex4-1}
\usepackage{graphicx}
\usepackage{dcolumn}
\usepackage{bm}
\usepackage{color}
\begin{document}

\title{Whispering gallery modes in hexagonal microcavities fabricated by crystal growth}
\author{Hiroshi Kudo}
\author{Ryo Suzuki}
\author{Takasumi Tanabe}  \email{takasumi@elec.keio.ac.jp}
\affiliation{Department of Electronics and Electrical Engineering, Faculty of Science and Technology,
Keio University, 3-14-1 Hiyoshi Kohoku-ku, Yokohama, 223-8522, Japan}

\date{\today}
\begin{abstract}
We investigated the whispering gallery modes of cavities with a hexagonal cross-section. We found two different modes, namely perturbed and quasi-WGMs, of which the former exhibits the higher $Q$ when the corner radius is large.  We studied the dependence of $Q$ on the curvature radius of the polygonal cavities and found that the coupling between the two modes determines the $Q$ of the cavity.  In addition we fabricated a cavity by employing laser heated pedestal growth and demonstrated a high $Q$.
\end{abstract}
\maketitle

\section{Introduction}
Optical microcavities such as photonic crystal nanocavities \cite{tanabe, noda}, microring resonators \cite{lipson}, toroidal microcavities \cite{kippenberg_nature} and crystalline whispering gallery mode (WGM) cavities \cite{crystalline} are attractive platforms for studying the optical linear and nonlinear properties of light \cite{kippenberg_sicence, ref19}.  These cavities have been employed in various studies including work on slow light generation \cite{tanabe, notomi_nature_phot, Vlasov}, an ultra-narrow linewidth laser \cite{maleki_pra}, an optoelectronic oscillator \cite{ref24, tanabe_pin}, nonlinear switching \cite{tanabe2, lipson}, frequency-comb generation \cite{kippenberg_sicence, maleki}, opto-mechanics \cite{kippenberg_optomechanics}, frequency conversion \cite{inoue, tanabe_prl}, and nanoparticle sensing \cite{vahala, vollmer}.  WGM cavities, which use internal reflection for light confinement, are particularly attractive because they can provide a very high quality factor ($Q$).  Recent progress on the $Q$ factor in such cavities has been noteworthy. In particular, cavities made of $\mathrm{SiO_{2}}$ such as toroidal \cite{ref21} and bottle \cite{ref2} cavities exhibit a very high-$Q$ exceeding $10^9$ because $\mathrm{SiO_{2}}$ is a low-loss material and the laser reflow process is well established.  On the other hand, WGM cavities made of crystal materials also exhibit an ultrahigh-$Q$, because their absorption loss is even lower than that of $\mathrm{SiO_2}$ \cite{maleki_pra}.  Crystalline materials are also attractive because they have various unique properties.  The large $\chi^{(2)}$ nonlinearity is essential for making building blocks for optoelectronic modulators \cite{ref24}.  It is transparent at infrared wavelengths, which is an interesting regime for gas sensing \cite{kippen_comb_ir}.  It also has a large Young's modulus, which is attractive for opto-mechanics research \cite{Kipp_mecha}.  With the development of ultra-precise machining, the properties of micro-resonators made of crystal materials have improved, and they are now good candidates especially as a platform on which to study nonlinear optics.

However, a high-$Q$ crystalline WGM cavity is still difficult to fabricate because the material is hard and frangible.  In particular, the fabrication of a small cavity remains a challenge.  Previously, we reported on the fabrication of sapphire ($\mathrm{Al_2 O_3}$) WGM cavities by employing a modified laser heated pedestal growth (LHPG) method to overcome this problem \cite{kudo}.  The LHPG method was originally developed for manufacturing fiber lasers \cite{ref8, ref9}.  It has been applied to various materials such as $\mathrm{Al_{2}O_{3}}$, $\mathrm{LiNbO_{3}}$ and YAG \cite{ref10}.  In contrast to earlier work, in which researchers tried to fabricate crystal rods with a thinner and uniform radius, we used this method to fabricate a rod with a locally modulated radius to enable light confinement in the longitudinal direction and excite the WGM.  Changing the growth rate enabled us to fabricate a swollen region, in which we obtained a WGM with a $Q$ of $1.6\times 10^4$.  In our work, we demonstrate the fabrication of sapphire WGM cavities that have a hexagonal cross-section but with round edges.  The controllability of the cross-section (the sharpness of the edges) is the key to controlling the $Q$ factor of such a cavity.

In this paper, we analyze in detail various WG-like modes excited in hexagonal cavities with different corner curvatures.  In particular, we study a quasi-WGM and a newly discovered perturbed-WGM.  To the best of our knowledge, this is the first time a perturbed-WGM has been characterized.  By understanding the property of the perturbed-WGM, and investigating the coupling between these two modes, we reveal the limit of the $Q$ of the quasi-WGM and study a possible way of improving the value.  We also demonstrate the fabrication and optical measurement of a sapphire hexagonal WGM cavity that supports such a mode.  We show that a high $Q$ cavity is obtained with a cavity with round corners, whose growth is carefully controlled to obtain good crystal quality.

The polygonal cavity analysis described in this paper is useful for cavities fabricated with LHPG, moreover it provides information needed to understand the optical properties of a polygonal toroidal microcavity \cite{ref22}, and microcavity lasers made of ZnO \cite{ref3, ref17}, $\mathrm{InGaAs}$ \cite{ref4}, $\mathrm{GaAs}$ \cite{ref27}, $\mathrm{GaN}$ \cite{ref28}, and $\mathrm{In_{2}O_{3}}$ \cite{ref29}, that are fabricated with the vapor phase transport method.  These microlasers have recently attracted great interest.  It is known that these microcavities have a polygonal cross-section as a result of their crystal structure, and various experimental and numerical studies has been conducted to understand the mode property of such cavities, especially in ZnO.  For example, Wiersing \cite{ref12}, Nobis and Grundmann \cite{ref5}, and Huang \cite{ref16} have analyzed the features of the modes in this unique cavity with triangular, hexagonal and circular cross-sections.  Crystals were grown experimentally on ZnO with different deposition temperatures.  This enabled the growth of ZnO micro wires with different cross-sections, including hexagonal, dodecagonal, and circular \cite{ref14}.   Dietrich \textit{et.~al} tried to improve the $Q$ value by rounding the corners of polygons \cite{ref26}.  However, the reported $Q$ of these microresonators remains on the order of several thousand, even though the excited modes, which they call a ``quasi-WGM'', seem to satisfy the total internal reflection in a ray-optics framework \cite{ref30, ref25}. Since we have a same problem in our sapphire hexagonal cavity, the purpose of the first half of this paper is to provide a clear understanding of the limiting $Q$ of this mode in a polygonal cavity.

This paper is organized as follows.  In Sec.~\ref{sec2} we introduce the structure of the cavity we fabricated for investigation.  We investigate the properties of the perturbed-WGM and quasi-WGM, paying particular attention to the influence of the coupling between these two modes.  Then in Sec.~III, we describe the fabrication and optical measurement of this cavity.  We conclude the paper in Sec.~IV.

\section{Analysis of hexagonal cavity}\label{sec2}
\subsection{Shape of microcavity fabricated by LHPG}
\begin{figure}[htbp]
 \centering
 \includegraphics[width=3.2in]{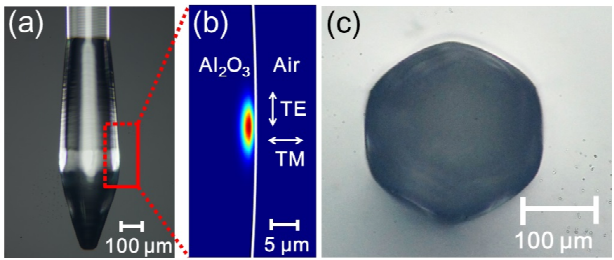}
 \caption{(color online) (a) Side view of the fabricated sapphire WGM microcavity.  (b) TM-mode profile calculated with the finite-element method (Comsol Multiphysics 4.3). (c) Cross-section of the fabricated sapphire WGM microcavity.}
 \label{fig:16}
\end{figure}

Before analyzing a hexagonal cavity with round corners, we show side and cross-sectional views of our hexagonal sapphire cavity fabricated by LHPG in Fig.~\ref{fig:16}. We describe the fabrication and optical measurement of this cavity in detail in Sec.~\ref{sec3}.
As shown in Fig.~\ref{fig:16}(a) the radius of the grown rod is locally modulated.  The light is confined in an area where the circumference is the largest, as shown in Fig.~\ref{fig:16}(b).  The typical mode volume is $1.28 \times 10^{-2}~\mathrm{mm^{3}}$.  Figure~\ref{fig:16}(c) shows an optical microscope image of the cross-section of this cavity.  The shape is hexagonal as a result of the crystal structure of the sapphire.  However, the corners of the cavity are not sharp but round, and their curvature radius can be controlled by changing the growth condition.  This cavity is on interest of mode analysis.
  \subsection{Definition of modes}
\begin{figure}[htbp]
 \centering
 \includegraphics[width=3in]{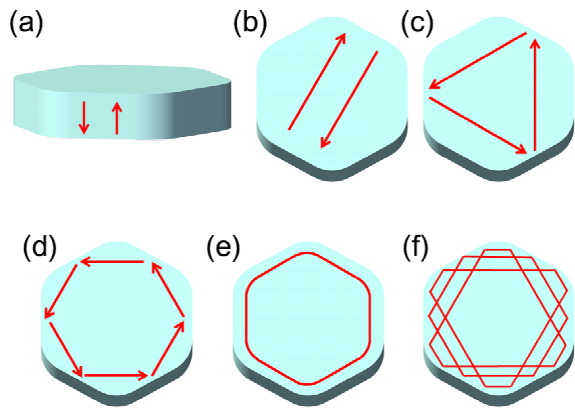}
 \caption{(color online) Illustrations of ray-optics descriptions of possible resonant cavity modes in hexagonal geometry. (a) and (b) are the Fabry-Perot modes.  (c) and (d) Quasi-WGM.  (e) Perturbed WGM.  (e) Higher order quasi-WGM.}
 \label{fig:0}
\end{figure}

Figure~\ref{fig:0} shows the ray-optics description of the possible modes in a hexagonal cavity with slightly rounded corners. Fabry-Perot (FP) modes (Fig.~\ref{fig:0}(a) and (b)) are excited by reflection at the opposing disk facets or at the hexagonal facets \cite{fabry, ref6}.  Figure~\ref{fig:0}(c) and (d) show ``quasi-WGMs'' in which light is reflected at the side of the polygon.  Since these modes satisfy the total internal reflection in a ray-optics framework, the quasi-WGM appears to exhibit an ultrahigh-$Q$.  Although, the quasi-WGM has been studied in detail in previous work \cite{ref30, ref25}, we will show below that the discovery of a different WGM, which is shown in Fig.~\ref{fig:0}(e), is essential if we are to understand the low-$Q$ of the quasi-WGM.  We call this mode a ``perturbed-WGM'' \cite{ref22}, because this is an ultrahigh-$Q$ WGM when the cavity is round.  A complete understanding of these modes is the main aim of this paper.  In addition, we show that a different mode, shown in Fig.~\ref{fig:0}(f), which has usually been categorized as a quasi-WGM, originates from a higher order WGM.

\begin{figure}[htbp]
\centering
\includegraphics[width=2.5in]{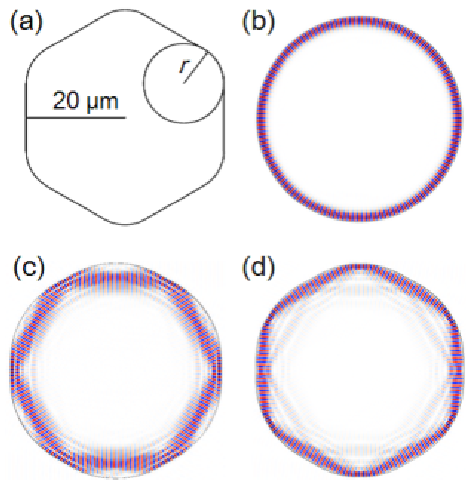}
 \caption{(color online) (a)~Model used for FDTD calculations.  The refractive index of the cavity is 1.74. The cavity radius is $20~\mathrm{\mu}$m and the edge radius is $r$. (b)~Mode profile of a WGM when the shape is a perfect circle ($r=\mathrm{20}~\mathrm{\mu m}$).  (c)~Mode profile of quasi-WGM when $r=\mathrm{16.1}~\mathrm{\mu m}$.  (d)~Mode profile of a perturbed-WGM when $r=\mathrm{16.1}~\mathrm{\mu m}$.}
 \label{fig:1}
\end{figure}

To investigate these modes in detail we perform a two-dimensional finite-difference time-domain (FDTD) calculation.  Figure~\ref{fig:1}(a) shows the structure that we used for the calculation.  It is a hexagonal cavity with round corners that have curvature radiuses $r$.

When the microcavity is a perfect circle, an ideal WGM is excited as shown in Fig.~\ref{fig:1}(b).  Next we change $r$ and make the cavity slightly polygonal.  Then we obtain an optical mode as shown in Fig.~\ref{fig:1}(c).  This is the quasi-WGM shown in Fig.~\ref{fig:0}(d).  The light propagates while being reflected at the side of the cavity.  This simple picture leads us to expect that the resonant wavelength and the $Q$ of the quasi-WGM do not depend strongly on the edge radius $r$.

In contrast, we obtained a different mode in a hexagonal cavity for the same geometry as shown in Fig.~\ref{fig:1}(d).  This is the ``perturbed-WGM''.  The original circular WGM is perturbed as a result of the modulation of the structure from circular to hexagonal.  But it retains the WGM characteristics, where the light propagates close to the surface.  Note that if we look very carefully, we find that the light of this mode propagates close to the surface at the corner, but it propagates slightly inwards at the side \cite{ref22}.
In contrast to the quasi-WGM, with the perturbed-WGM we can straightforwardly predict that the resonant wavelength and $Q$ value change sensitively according to the corner radius $r$.

Since this mode exhibits an ultrahigh-$Q$ when the corner is round, we expect the $Q$ of the perturbed-WGM to be higher than that of the quasi-WGM when $r$ is large.  However, the $r$ dependence of the $Q$ is larger for the perturbed-WGM, therefore, the $Q$ of the quasi-WGM should become higher when the corner is sharp and the cavity is hexagonal.  But, this simple description is insufficient as we discuss in the following sections.

\subsection{Mode analysis}
\subsubsection{Resonant modes in hexagonal cavity}
\begin{figure}[htbp]
 \centering
 \includegraphics[width=3in]{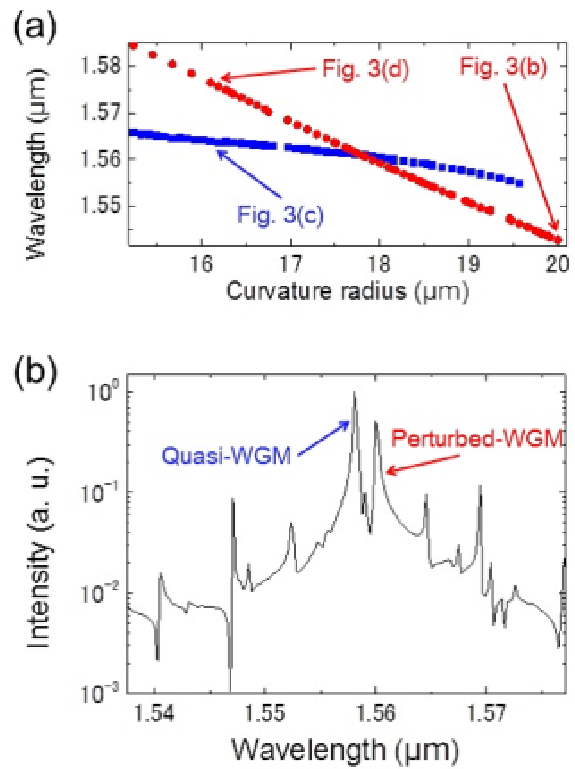}
 \caption{(color online) (a)~Resonance wavelength of perturbed-WGM and quasi-WGM in relation to the curvature radius $r$.  (b)~Spectrum of the excited modes of a hexagonal cavity when $r=17.8~\mathrm{\mu m}$.}
 \label{fig:2}
\end{figure}
Figure~\ref{fig:2}(a) shows the resonance wavelength of the perturbed and quasi WGMs with respect to the $r$ calculated using the FDTD method.  As expected, the resonant wavelength changes much sensitive to the $r$ for the perturbed-WGM than for the quasi-WGM.  But most importantly, we know from Fig.~\ref{fig:2}(a) that the WGM of the circular cavity shown in Fig.~\ref{fig:1}(b) is on the same slope as the perturbed WGM in Fig.~\ref{fig:1}(d).  This indicates that the mode shown in Fig.~\ref{fig:1}(d) is indeed a ``perturbed'' WGM, and the quasi-WGM originates from a different mode from an ultrahigh-$Q$ WGM when the cavity is circular.

\begin{figure}[htbp]
\centering
\includegraphics[width=3in]{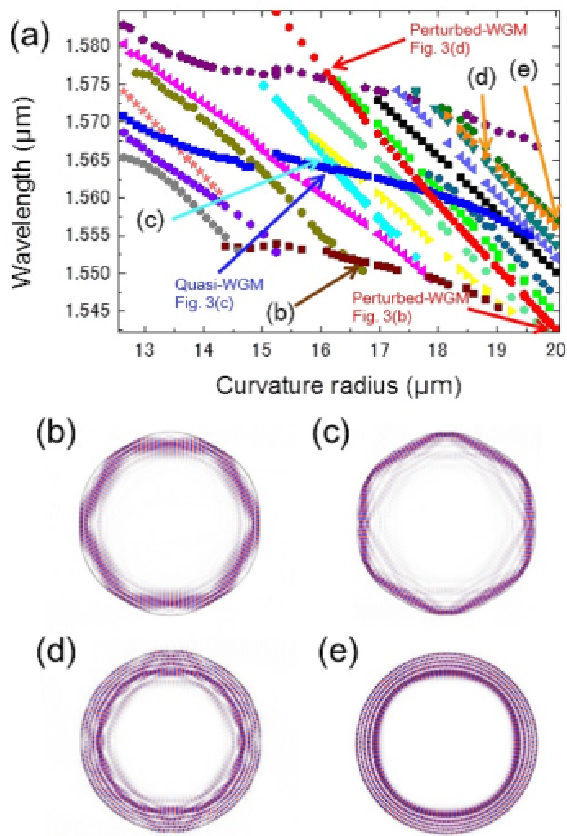}
\caption{(color online) (a)~Resonant modes of hexagonal cavities with different curvature radiuses $r$.  (b-e)~Mode profiles of the cavity.  The corresponding resonance points are shown in (a).}
\label{fig:3}
\end{figure}
Next, we investigate the resonance in more detail.  Figure~\ref{fig:2}(b) shows the resonant spectrum when $r = 17.8~\mathrm{\mu m}$.  This spectrum shows that not only two modes, but also various other different modes are excited in this hexagonal cavity.  To explain the property of the hexagonal cavity, we need to understand these modes.  Since it is difficult to characterize the property of this number of modes from this single spectrum, we need to perform calculations with different $r$.

The resonant peaks for cavities with different $r$ values are shown in Fig.~\ref{fig:3}(a). The plots for the quasi and perturbed WGMs shown in Figs.~\ref{fig:1}(c) and (d) are indicated in the panel.  However, a number of different peaks appear in addition to these two modes.  Although the behavior of these modes appears very complicated, they can be classified into two different groups.  We clearly obtained two different slopes.

Figure~\ref{fig:3}(b) shows a quasi-WGM but with a different longitudinal mode number.  Similar resonance also appears at an 11.6 nm longer wavelength, where the interval is the free-spectral range (FSR) of this mode.  Indeed, we obtained a similar FSR value from a simple ray-optics estimation.  Similarly, Fig.~\ref{fig:3}(c) shows a perturbed-WGM with a different longitudinal mode number.  The FSR for the perturbed-WGM is 11.3~nm.

However, Fig.~\ref{fig:3}(c) shows that there are a number of different modes with a similar $r$ dependence to that of the perturbed-WGM in the hexagonal cavity. One of the modes is shown in Fig.~\ref{fig:3}(d). In this mode, the light propagates not only at the surface of the cavity, but also at the inner part of the structure.  This propagation occurs because the light reflects at the cavity wall with a large incident angle.  The light propagation in this mode is described in terms of ray-optics in Fig.~\ref{fig:0}(f), where the mode was originally classified as a quasi-WGM. However, the $r$ dependence of this mode in Fig.~\ref{fig:3}(b) suggests that its behavior is similar to that of the perturbed-WGM.  Indeed, when we examine the profile at $r=20~\mathrm{\mu m}$ for this mode, namely when the cavity is circular, we observe a higher order WGM as shown in Fig.~\ref{fig:3}(e).  Hence, we can conclude that the mode shown in Fig.5(d) is the perturbed mode of the higher order WGM.

It is important to understand this picture, because it indicates the difficulty of obtaining a high-$Q$ for the mode shown in Fig.~\ref{fig:3}(e), even though the ray-optics image in Fig.~\ref{fig:0}(f) gives the impression that this mode may exhibit an ultrahigh $Q$ even when the corner is sharp.

\subsubsection{Cavity $Q$ of perturbed and quasi WGMs}
\begin{figure}[tbhp]
\centering
\includegraphics[width=3in]{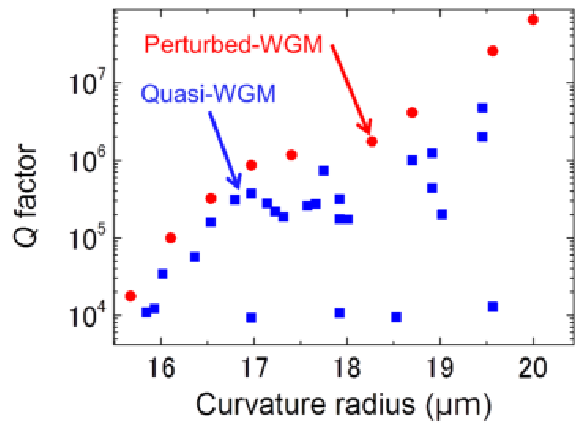}
\caption{(color online) The $Q$ of perturbed and quasi WGMs for hexagonal cavity with different curvature radius $r$ values.}
\label{fig:4}
\end{figure}

Since the quasi and perturbed WGMs (Fig.~\ref{fig:1}(c) and (d)) are the fundamental modes of this hexagonal cavity, we now focus on these two modes and analyze the $Q$.  Figure~\ref{fig:4} shows the $Q$s of these two modes with different $r$ values.  Contrary to our initial expectation, the quasi-WGM exhibits a lower $Q$ when the cavity is polygonal, although Fig.~\ref{fig:0}(d) suggests that the $Q$ is only slightly dependent on $r$.  

To understand this behavior, we pay close attention to the wavelength crossing between the quasi-WGM and the perturbed-WGM.

\subsubsection{Coupling between modes}
\begin{figure}[htbp]
\centering
\includegraphics[width=3in]{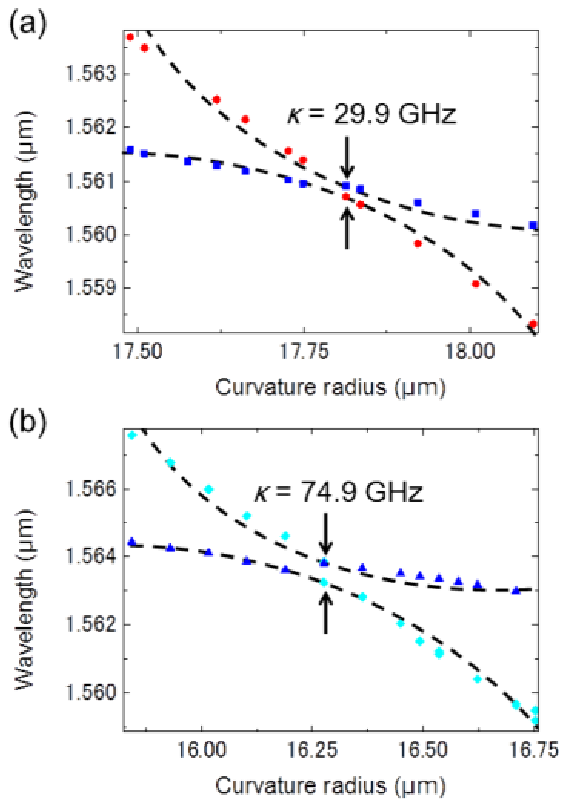}
\caption{(color online)  Wavelength crossing between perturbed and quasi WGMs.}
\label{fig:6}
\end{figure}
Figure~\ref{fig:6}(a) and (b) show enhanced views of the crossing points between the perturbed and quasi-WGMs in Fig.~\ref{fig:3}(a).  The figures show that these two modes exhibit anti-crossing behavior, which indicates the existence of strong mode coupling between those two modes.  We can derive the coupling coefficient $\kappa$ from the spectrum splitting width, and we obtained $\kappa=2.99 \times 10^{10}~\mathrm{s}^{-1}$ for Fig.~\ref{fig:6}(a) and $\kappa=7.49 \times 10^{10}~\mathrm{s}^{-1}$ for Fig.~\ref{fig:6}(b).  This explains the low $Q$ property of the quasi-WGM when the corners are sharp.

According to the ray-optics description shown in Fig.~\ref{fig:0}(d), the $Q$ of the quasi-WGM does not depend on the edge radius $r$ of the hexagonal cavity and should exhibit an ultrahigh $Q$.  However, the coupling between the quasi- and perturbed-WGMs needs to be taken into account.  This coupling is particularly important when $r$ is small.  Coupling between modes, for example $\kappa=2.99 \times 10^{10}~\mathrm{s^{-1}}$, which corresponds to a $Q$ of about $3 \times 10^3$, means that the light energy of the high-$Q$ quasi-WGM transfers to a low-$Q$ perturbed-WGM and then easily couples to the out-of- cavity radiation.  Therefore, the $Q$ of the quasi-WGM is not dependent on the $Q$ of the perturbed-WGM and the values of those two modes become very close.  Since the $Q$ value of the perturbed-WGM is low when the corner is sharp, the $Q$ value of the quasi-WGM also decreases.  The very low $Q$ plots ($\sim 10^4$) of the quasi-WGM in Fig.~\ref{fig:4} result from coupling with higher order perturbed-WGMs that have a low $Q$.

A numerical analysis revealed that the $Q$ of quasi-WGM is fundamentally dependent on the $Q$ of the perturbed-WGM.  Since the $Q$ of the perturbed-WGM is low when the corner radius $r$ is small, it is not easy to obtain a high-$Q$ cavity even when we excite the quasi-WGM.  Therefore, the only method with the potential to provide a high $Q$ in such a crystalline cavity is to make the shape as circular as possible.  In Sec.~\ref{sec3}, we show methods for fabricating circular sapphire WGM cavities by using LHPG to achieve a high $Q$.

\subsection{Optimal size of polygonal cavity}
In this section, we discuss the optimal radius of the hexagonal cavity.  We focus on the perturbed-WGM in a hexagonal cavity with a radius $R$ and a curvature radius $r~=~0.8R$.  It is known that the $Q$ of a circular WGM cavity increases linearly as the radius increases \cite{crystalline}.  However, this characteristic has not been studied for a polygonal WGM cavity.

\begin{figure}[htbp]
\centering
\includegraphics[width=2.5in]{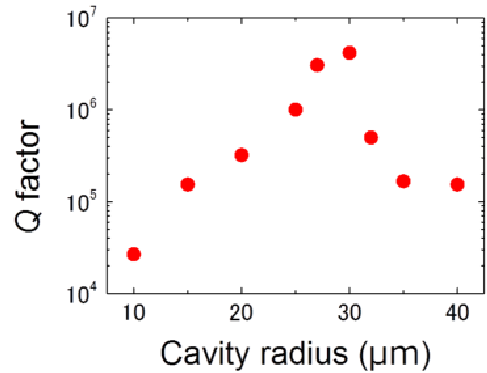}
\caption{(color online) $Q$ factor of perturbed-WGM as a function of the radius $R$ of hexagonal cavities with a curvature radius $r=0.8R$.}
\label{fig:diameter}
\end{figure}
Figure~\ref{fig:diameter}(a) shows the $Q$ factor for the perturbed-WGM for hexagonal cavities with different cavity radius $r$.  It clearly shows that there is an optimal point at $r=30~\mathrm{\mu m}$.  The $Q$ value decreases when $d$ is small, because it is difficult for the light in the cavity to satisfy the total internal reflection condition when the cavity is small.  Moreover, unlike a circular cavity, the $Q$ value decreases when $r~>~30~\mathrm{\mu m}$, which is difficult to explain solely in terms of total internal reflection.  A careful investigation revealed that mode coupling plays an important role.  In a large cavity, the mode spacing of the lowest order perturbed-WGM and that of the higher order perturbed-WGM are both very small, which results in an easy overlap between these modes.  Since higher order-WGMs have a much lower $Q$, the $Q$ of the lowest-order perturbed-WGM also decreases for a large cavity.  Therefore, there is an optimum radius as regards obtaining a high-$Q$ in a hexagonal cavity.

\section{Experiment}\label{sec3}
\subsection{Experimental setup}
\begin{figure}[htbp]
\centering
\includegraphics[width=2.5in]{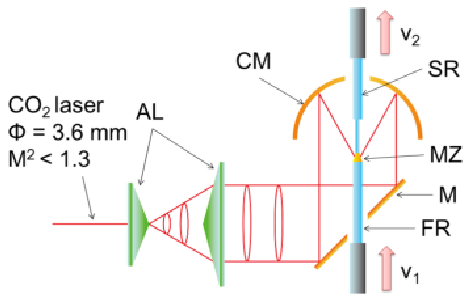}
\caption{Schematic illustration of experimental setup. AL: ZnSe axicon lens, FR: feed rod, SR: seed rod, M: gold flat mirror with a hole at the center, CM: concave gold mirror (curvature = 100~mm) with a hole at the center, MZ: molten zone. A $\mathrm{CO_{2}}$ laser with diameter $\phi$ and beam parameter product $M^2$ is used to heat the crystal rods.}
\label{fig:10}
\end{figure}

Figure~\ref{fig:10} shows the experimental setup for the LHPG method.  The input $\mathrm{CO_{2}}$ laser beam  (Coherent DIAMOND C-70) is formed into a doughnut shape by using a pair of axicon lenses.  This minimizes the beam that hits the feed rod before it is focused with a concave mirror, and it makes the beam focusing more efficient.  The $\mathrm{CO_{2}}$ laser beam is focused on the feed rod by using a concave mirror with a radius of 50~mm to heat the rod and form a molten zone.  The laser beam spot size is about $\mathrm{12.2~\mu m}$, and the power density is $8.58 \times 10^7~\mathrm{W/m^{2}}$ when the $\mathrm{CO_{2}}$ laser power is 4.0~W.  Both the seed and feed rods are fixed to a linear translation stage (Oriental Motor ASM46MA).  The radius of the sapphire single crystal rods is $213~\mathrm{\mu m}$ for both the feed and seed rods.

Then a crystal fiber is grown by pulling the feed rod upwards at a speed of ${v_{1}}$, while the seed rod is moved in the same direction with a speed of $v_2$ ($v_2 > v_1$).  In LHPG, we can control the radius of the grown crystal rods by changing the velocity rate whose relationship is given by
\begin{equation}
R_1=R_2\left(\frac{v_2}{v_1}\right)^{\frac{1}{2}},
\end{equation}
where ${R_1}$ and ${R_{2}}$ are the radiuses of the grown and feed rods, respectively.  LHPG was originally developed to create crystal rods with a uniform radius. In contrast, by changing the velocity rate we can create a bulge with which we can confine light in the longitudinal direction and excite the WGM.

In the experiment, we changed the velocity rate from 1/6 to 1/3.  This corresponds to calculated $R_1$ values of $\mathrm{87~\mu m}$ and $\mathrm{123~\mu m}$, when the feed rod radius $R_2$ is $213~\mathrm{\mu m}$.  First, we started with standard LHPG. $v_1$ was set at $\mathrm{2~\mu m/s}$ and ${v_2}$ was set at $\mathrm{12~\mu m/s}$.  In brief, the $v_1/v_2$ was 1/6 and the grown rod was fabricated with a radius of $87~\mathrm{\mu m}$. Next, we reduced $v_2$ from $12~\mathrm{\mu m/s}$ to $6~\mathrm{\mu m/s}$ and set the $v_1/v_2$ ratio at 1/3 to fabricate a WGM cavity.  The grown radius of the WGM cavity was then larger at $123~\mathrm{\mu m}$. Finally, we stopped the seed rod and pulled the feed rod downward.

A side view and cross-section of the fabricated cavity are shown in Figs.~\ref{fig:16}(a) and (c).  The cross-section of this cavity is hexagonal as a result of the crystalline structure of the sapphire.  As discussed earlier in this paper, it is essential to develop a method for controlling the curvature of the edge radius of this cavity in order to increase the $Q$.

\subsection{Control of cross-section geometry}
\begin{figure}[htbp]
\centering
\includegraphics[width=3.0in]{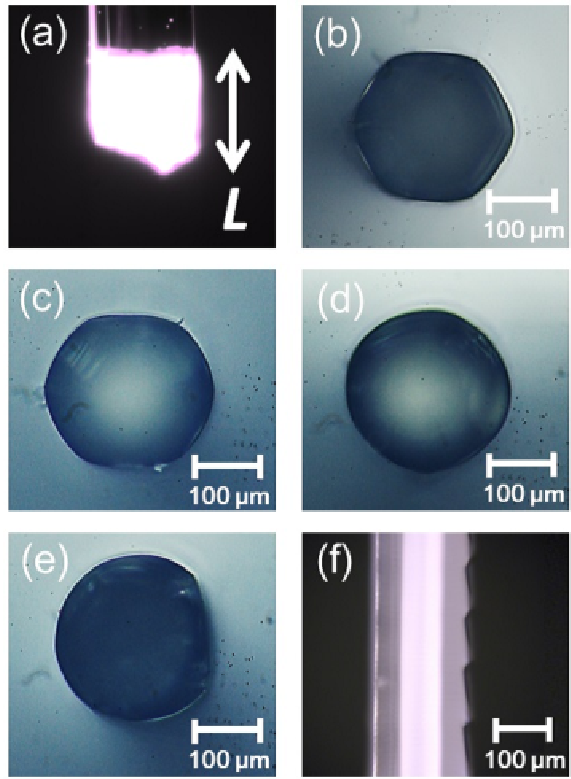}
\caption{(color online) (a)~Optical microscope image of a seed rod during the preheating process.  $L$ is the length of the preheated region.  (b)-(e) Optical microscope image of a cross-section of the fabricated sapphire cavity with different $L$ values. (b) $L=0.10~\mathrm{mm}$, (c) $L=0.70~\mathrm{mm}$, (d) $L=0.78~\mathrm{mm}$, and (e) $L=0.80~\mathrm{mm}$.  (f)~Side view of the microcavity fabricated when $L$= 0.80~mm.}
\label{fig:22}
\end{figure}

We introduce a pre-heating method to make the cavity round.  The seed rod is heated with a $\mathrm{CO_2}$ laser at a power of 8~W prior to the crystal growth.  Figure~\ref{fig:22}(a) shows the $213~\mathrm{\mu m}$ radius pre-heating seed rod , where $L$ is the heating length.  We changed $L$, and observed the cross-sectional shape of the fabricated WGM cavity.  Figure~\ref{fig:22}(b)--(e) show the cross-section of the fabricated sapphire cavity with different $L$ values.  When $L$ is larger, the cross-sectional shape becomes close to a circle, and we obtained an optimal $L$ value of 0.78~mm.

This experimental result clearly shows that the preheating method is a promising way to control the cross-sectional shape of the cavity.  However, when $L$ is 0.80~mm or larger, the cross-section deforms as shown in Fig.~\ref{fig:22}(e).  In particular, the side view of the cavity (Fig.~\ref{fig:22}(f)) shows that the reduction of the crystal quality might be significant.  Therefore, we need to employ the X-ray diffraction (XRD) technique to investigate the crystal quality of the fabricated cavity.

\subsection{X-ray analysis}
\begin{figure}[htbp]
\centering
\includegraphics[width=2.5in]{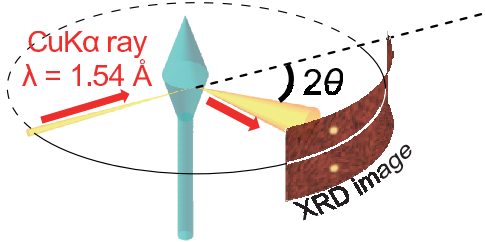}
\caption{(color online) Experimental setup for XRD measurement.}
\label{fig:xrd}
\end{figure}

Since we expected a tradeoff between the shape and the crystalline quality of the cavity, we conducted XRD spectrum measurements for cavities with different shapes.  Because we are interested in the crystal quality at the surface of the cavity where light propagates, we use the $2\theta$--$\theta$ method, which has an x-ray penetration depth of a few $\mathrm{\mu m}$ from the sample surface.  The experimental setup is shown in Fig.~\ref{fig:xrd}.

\begin{figure}[htbp]
\centering
\includegraphics[width=3.2in]{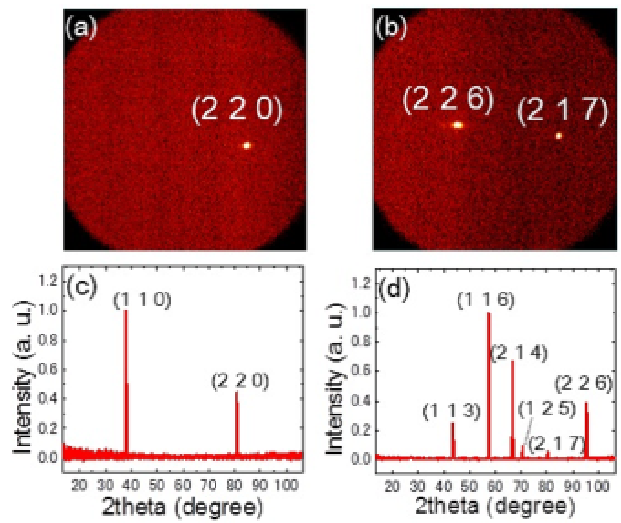}
\caption{(color online) (a)~2D image of XRD of a hexagonal cavity.  The $2\theta$ range in this image is from $73.3^\circ$ to $106.7^\circ$.  The corresponding Miller indices of $\mathrm{Al_{2}O_{3}}$ are shown in the panel.  (b)~Same as (a) but with a circular cavity.  (c) XRD spectrum of the hexagonal cavity.  (d) Same as (c) but with a circular cavity.}
\label{fig:20}
\end{figure}

Figures~\ref{fig:20}(a) and (b) show 2D images of XRD measurements of cavities with hexagonal (Fig.~\ref{fig:16}(c)) and circular cross-sections. (Note: The circular cavity used for this experiment is not the same as that shown in Fig.~\ref{fig:22}(d), but it has an almost identical shape.)  The x-ray was incident perpendicular to the c-axis of the crystal.  Most importantly, in these images we observe bright spots, which indicate that the crystal quality is maintained.  Note that Debye-Scherrer rings should appear if the crystallinity is low.

Figures~\ref{fig:20}(c) and (d) are the XRD spectra of these cavities.  We observed diffraction peaks that originated from $\mathrm{Al_2 O_3}$, but there is no clear evidence of different alumina compounds.  This also shows that relatively good crystal quality is maintained even for a circular cavity.  However, we found that the spectrum peaks diffracted owing to the high-order planes as seen in Fig.~\ref{fig:20}(b).  This indicates that in some part of the cavity, the c-axis of the crystal is tilted.   We must be careful if we want to use the nonlinear susceptibility of this material, which is a tensor.  However, for other applications, this effect is limited, although some randomness is inherent in the crystal structure.

\subsection{Optical measurement}
\begin{figure}[htbp]
\centering
\includegraphics[width=3in]{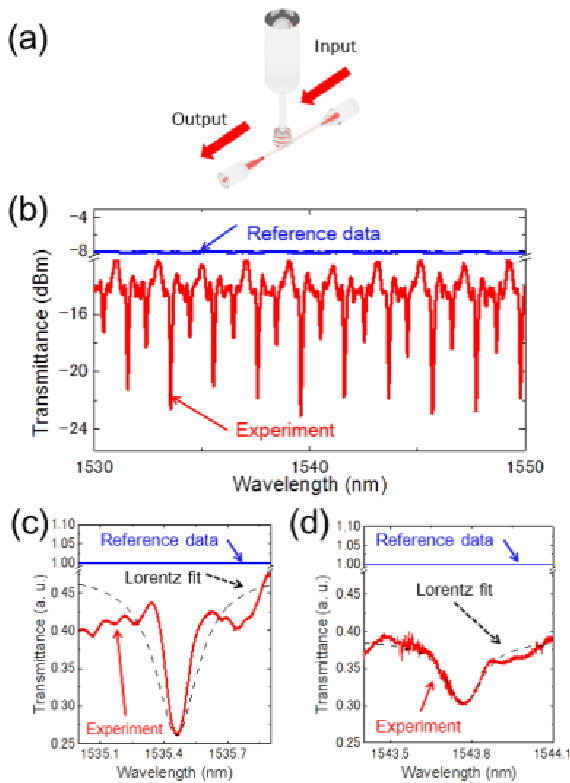}
\caption{(color online) (a)~Schematic illustration of optical experimental setup using a conventional tapered fiber.  (b) Transmission spectrum of a circular cavity fabricated using preheating. (c)~Enlarged image of (b). The dashed line is a Lorentzian fit for the experimental data, whose width gives a $Q$ of $\mathrm{1.6 \times 10^{4}}$.  (d)~Transmission spectrum of the hexagonal cavity without preheating. It has a $Q$ of $\mathrm{8.5 \times 10^{3}}$.}
\label{fig:14}
\end{figure}

We measured the transmittance spectrum of the sapphire microcavity in Fig.~\ref{fig:14}(a) by coupling light with an optical tapered fiber setup.  We can control the coupling between the resonator and the waveguide in order to measure the cavity $Q$.  The transmittance spectrum of a circular sapphire WGM cavity is shown in Fig.~\ref{fig:14}(b).  We observe clear resonance with an equal free-spectral range.  Figure~\ref{fig:14}(c) is an enlarged view of Fig.~\ref{fig:14}(b).  By fitting the spectrum with a Lorentzian shape, we obtained a $Q$ of $1.6 \times 10^{4}$.  This value is higher than that obtained from the resonant spectrum of a hexagonal cavity shown in Fig.~\ref{fig:14}(d),  which has a $Q$ of $\mathrm{8.5 \times 10^{3}}$.  As expected, with a numerical analysis, we obtained a higher $Q$ for a circular cavity than for a hexagonal cavity.

Although theory and experiment both predict a higher $Q$ for a circular cavity, and are in a good agreement, the absolute $Q$ values are far apart.  To improve the experimental $Q$, we discuss the effect of surface roughness and perform some experiments to improve the smoothness.

\subsection{Surface roughness}
One reason for the low-$Q$ in the experiment is the relatively large radius of the WGM cavity, which results in a lower $Q$ as described in Sec.~\ref{sec2}.  However, different factors can influence the cavity $Q$.  Of these, surface roughness is clearly the most important factor in terms of impact on the cavity $Q$.  It is known that the $Q$ value is proportional to the square of the surface roughness \cite{crystalline}.  Here we briefly describe an experiment designed to improve the smoothness of the surface by changing the crystal growth velocity.

\begin{figure}[htbp]
\centering
\includegraphics[width=3in]{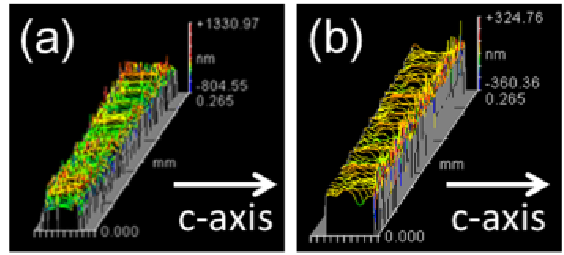}
\caption{(a) Surface plot of the same circular microcavity shown in Fig.~\ref{fig:14}(c).  The roughness is 65~nm.  (b)~Surface plot of a crystal rod grown with a velocity of $\mathrm{2~\mu m/s}$.  The surface roughness is 6~nm.}
\label{fig:15}
\end{figure}

We used a scanning white-light interferometer (Zygo New View TM6200) to examine the surface roughness of the side wall of the cavity.  Figure~\ref{fig:15}(a) is a 3D surface plot for a circular cavity with a $Q$ of $1.6 \times 10^{4}$.  We fabricated this cavity with a crystal growth speed of $\mathrm{12~\mu m/s}$.  The average surface roughness of the cavity surface was 65-nm RMS.  According to Ref.~\cite{crystalline}, we may be able to obtain a $Q$ on the order of $10^7$ if we can reduce the surface roughness to a few nm.

To fabricate a crystal rod with a smoother surface, we reduced the growth velocity to $2~\mathrm{\mu m/s}$.  Then we obtained a surface roughness of about 6~nm RMS, as shown in Fig.~\ref{fig:15}(b).  The result shows us that the growth velocity is the key to the surface roughness. 

Although optimization is still being studied, we should be able to fabricate a crystalline WGM cavity with a smoother surface, which may support an ultrahigh $Q$.  A different method of optimizing the surface might involve growing the crystal in a He gas environment \cite{atmosphere}.  Since various studies have been conducted using the LHPG method, there is still plenty of room left for the optimization of the surface smoothness.

\section{Conclusion}
We investigated excited modes in polygonal cavities in detail and found that a perturbed-WGM exhibits an ultrahigh $Q$ when the corner radius of the cavity is large.  We found that a number of lower and higher order perturbed-WGMs are present in a hexagonal cavity coupled with a quasi-WGM.  Although the quasi-WGM appears to exhibit an ultrahigh-$Q$ with ray-optics, mode coupling with the perturbed-WGM limits the $Q$ of this mode.  We revealed that we need to understand the coupling between the modes to explain the low $Q$ of the quasi-WGM when the cavity is hexagonal.  As a result, the only way to improve the $Q$ of the crystalline cavity (with both sapphire and ZnO cavities) is to make the corner round.  With this in mind, we showed pre-heating method that can control the cross-section of the fabricated cavity.  Although some degeneracy of the crystal was observed, this effect was limited.  Despite the experimental $Q$ of the current device being limited by the surface roughness, we showed that further optimization is possible.

Research using WGM cavities is active in various fields including micro-laser fabrication, quantum-optics, sensing, and signal processing.  However, the fabrication of a crystalline WGM remains difficult because of its fragility.  This study will contribute to the provision of an alternative method for fabricating very small crystalline WGMs by LHPG for use in these fields.

\section*{Acknowledgments}
Part of this work was supported by the Strategic Information and Communications R\&D Promotion Programme (SCOPE), a Grant-in-Aid for Scientific Research from the Ministry of Education, Culture, Sports, Science and Technology, Japan (KAKEN \#25600118), and the Keio University Next Generation Research Project Promotion Program.  The author would also like to thank Dr. Atsushi Yokoo from NTT Basic Research Laboratories, Prof. Yasuhiro Kakinuma from Keio University, and Prof. Yoichi Kamihara from Keio University, for valuable discussions on LHPG, precise machining and XRD analysis, respectively.

\end{document}